\documentclass{article}[margin=1in]
\usepackage[utf8]{inputenc}
\usepackage[margin=1in]{geometry}
\usepackage{authblk}
\usepackage{color}
\usepackage{url}
\usepackage[super]{natbib}
\usepackage{hyperref}
\usepackage[utf8]{inputenc}
\usepackage{graphicx}

\title{Computational and Systems Biology Advances to Enable Bioagent Agnostic Signatures}

\author[1]{\footnotesize Andy Lin \footnote{Equal Contribution} \footnote{Corresponding Author}}
\author[2]{Cameron Torres*}
\author[1]{Errett C. Hobbs}
\author[3]{Jaydeep Bardhan}
\author[2]{Stephen Aley}
\author[2]{Charles T. Spencer}
\author[1]{Karen L. Taylor}
\author[1,2,4]{Tony Chiang $\dagger$}

\affil[1]{National Security Directorate, Pacific Northwest National Laboratory, Seattle, WA 98109, USA}
\affil[2]{Department of Biological Sciences, University of Texas at El Paso, El Paso, Texas 79968 USA}
\affil[3]{Earth and Biological Sciences Directorate, Pacific Northwest National Laboratory, Seattle, WA 98109, USA}
\affil[4]{Department of Mathematics, University of Washington, Seattle 98102 USA}

\date{}

\begin{document}

\maketitle

\begin{abstract}
Enumerated threat agent lists have long driven biodefense priorities. The global SARS-CoV-2 pandemic demonstrated the limitations of searching for known threat agents as compared to a more agnostic approach. Recent technological advances are enabling agent-agnostic biodefense, especially through the integration of multi-modal observations of host-pathogen interactions directed by a human immunological model. Although well-developed technical assays exist for many aspects of human-pathogen interaction, the analytic methods and pipelines to combine and holistically interpret the results of such assays are immature and require further investments to exploit new technologies. In this manuscript, we discuss potential immunologically based bioagent-agnostic approaches and the computational tool gaps the community should prioritize filling.
\end{abstract}

\section{Introduction}
Historically, threat identification and characterization has centered on lists of specific agents known to cause severe harm to human, animal, or agricultural health, such as the Federal Select Agent Program Select Agent and Toxin list \cite{federal:select}, the WHO Prioritization List \cite{World:prioritizing}, and the National Institute of Allergy and Infectious Diseases (NIAID) Emerging Infectious Diseases and Pathogens List \cite{U.S.:NIAID}. List-based biodefense approaches focus on agents that have been part of state-sponsored biological weapons programs or are especially dangerous known natural threats. However, previous work has highlighted the limitations of framing biodefense strategies around defined lists \cite{leiser:beyond}. List-based approaches are ill-equipped to accommodate threats posed by emergent, re-emergent, or novel pathogens, as demonstrated by SARS-CoV-2. While there were warnings about coronaviruses with pandemic potential \cite{mahroum:covid19}, SARS-CoV-2 did not appear in any lists prior to the outbreak and caused significant loss of life and disrupted social order worldwide. Early, rapid detection and characterization of unknown pathogens is an integral component to a robust biodefense posture. In addition, characterizing how an unknown agent will likely affect human, animal, and plant health is a crucial requirement for biopreparedness and response.

Leiser et al. proposed a strategy to augment list-based approaches by characterizing threats based on how they affect human, animal, or plant hosts\cite{leiser:beyond}. Specifically, they advocated that the biodefense community should shift from a identification-based approach to a characterization-based one by developing bioagent-agnostic signatures (BASs), defined as measurable suites of biomarkers that accurately and reproducibly assess the impacts of infection or intoxication without a priori knowledge of an agent \cite{leiser:beyond}.

Retooling the US biodefense posture from a list-based approach to a dual list-based and BAS-based approach will require policy changes, technological improvements, and improvements in data analytics. Encouragingly, recent policy shifts have signaled how the United States government recognizes the need for new technologies to counter chemical and biological threats beyond the preexisting lists of anticipated pathogens and toxins\cite{seligman:new}. Despite this awareness, the biodefense community has only just begun to develop technologies for identifying useable BASs.

Enabling a BAS-approach to complement existing list-based approaches to biodefense would require additional investment. Specifically, development of BASs would work in two dimensions: 1) pathogen characterization/classification and 2) host response characterization. While the biodefense community has started to develop technologies to categorize and detect BASs, the community has not addressed the data science challenges that hinder this new approach. This perspective highlights some promising new immunological approaches that could be leveraged for BAS development, data science problems that the community must resolve to identify BASs, and possible ways forward.

\section{Threat Agnostic Detection Technologies}

\subsection{Opportunities to Infer Pathogenicity from Host Response}
Currently, biological agent detection relies on screening samples against existing databases (e.g., polymerase chain reaction (PCR), sequencing, or proteomics) or probing with antibodies specific for particular entities (e.g., lateral flow immunoassays (LFIAs) or enzyme-linked immunosorbent assays (ELISA)) \cite{Kaslow2014,Peruski2003}. Despite their reliability for diagnosis and epidemiology \cite{Klebes2022,Kang2021,Law2014}, they unfortunately lack agent-agnostic characteristics.

One promising direction for developing BASs would involve identifying signatures based on the host immune response, rather than a predetermined sequence.
Detection of foreign invaders by the host's immune system relies first on the generation of binding protein receptors capable of detecting the broadest possible range of epitopes, initially non-specific to any particular agent. After generation in the body, these proteins are subjected to negative selection that removes any self-reactive receptors. Ideally, whatever survives this selection process must recognize non-self, foreign antigens. 
Just as innate immune receptors identify classes of microbes through common structures, so too might an agent-agnostic detection approach discern structures not represented in human populations ({\it e.g.}, terminal $\alpha1-3$ galactosyl moieties in parasites \cite{Magnus2020}). This approach would create a new type of assay that would be a proxy for immune cells, that simultaneously interrogates numerous analytes and integrates data from across multiple platforms to detect BASs.

\subsection{Promising Biotechnology for Bioagent-Agnostic Detection}
Currently, no single screening model is suitable for BAS discoveries since they rely on highly specific molecules or searches against established references. Development of new technologies coupled with modifications to existing methods is needed to enable BAS. For example, creation of an artificial immune system could allow for the direct testing of unknown samples (either environmental or clinical) for the presence of a foreign entity to which the ``immune system on a chip'' responds. Technologies utilizing cell responses have begun these advancements for clinical use\cite{morse:clinical,zhou:cd8+}, yet they are still limited by the need to directly analyze selected immune factors determined from analysis of specific diseases rather than general factors indicative of a response to unknown pathogens. From the environmental sample perspective, the DARPA ``Friend or Foe'' program has developed technologies that use cell response to differentiate between pathogenic and non-pathogenic bacteria\cite{phillips:engineered}.

Recent modifications to flow cytometry, specifically the development of mass cytometry \cite{tanner:introduction,spitzer:mass}, provide another example of how modern technologies can advance the goal of threat-agnostic biodefense. While flow cytometry has been used for decades for cell analysis, its sensitivity for BAS is capped because of the limited number of  predefined cell surface markers. 
An expanded number of markers at a higher specificity is now possible with mass cytometry. It also allows for the simultaneous detection of immune cells and changes in their protein expression as biomarkers for threat detection \cite{Zhou2019,Chrdle2019}. This has the potential to be adapted to monitor a wide range of analytes by creating a ``multiplexed'' panel of biomarkers/BASs indicative of a threat. The future potential for measuring multiple 'omics (e.g., multi-omics) within a single cell \cite{Fulcher2022.05.17.492137} also provides a powerful tool to probe multiple cell types, such as an infected cell (e.g., lung epithelial) or an immune cell (e.g., CD8+ cytotoxic T-cells) for host response, which is important to capture pathogenic pathways that target specific cells or tissues (e.g., a respiratory pathogen may have little effect on a muscle cell).
These technologies provide insight unavailable a few decades prior. However, 
as we discuss in future sections, computational analysis such as batch corrections in cytometry and unsupervised learning for clustering heterogeneous populations of single cells remain challenging.

\subsection{Agnostic Technologies Require Collection of Multi-Omics Data}
\begin{figure}
    \centering
    \begin{tabular}{l}
    \includegraphics[width=6in]{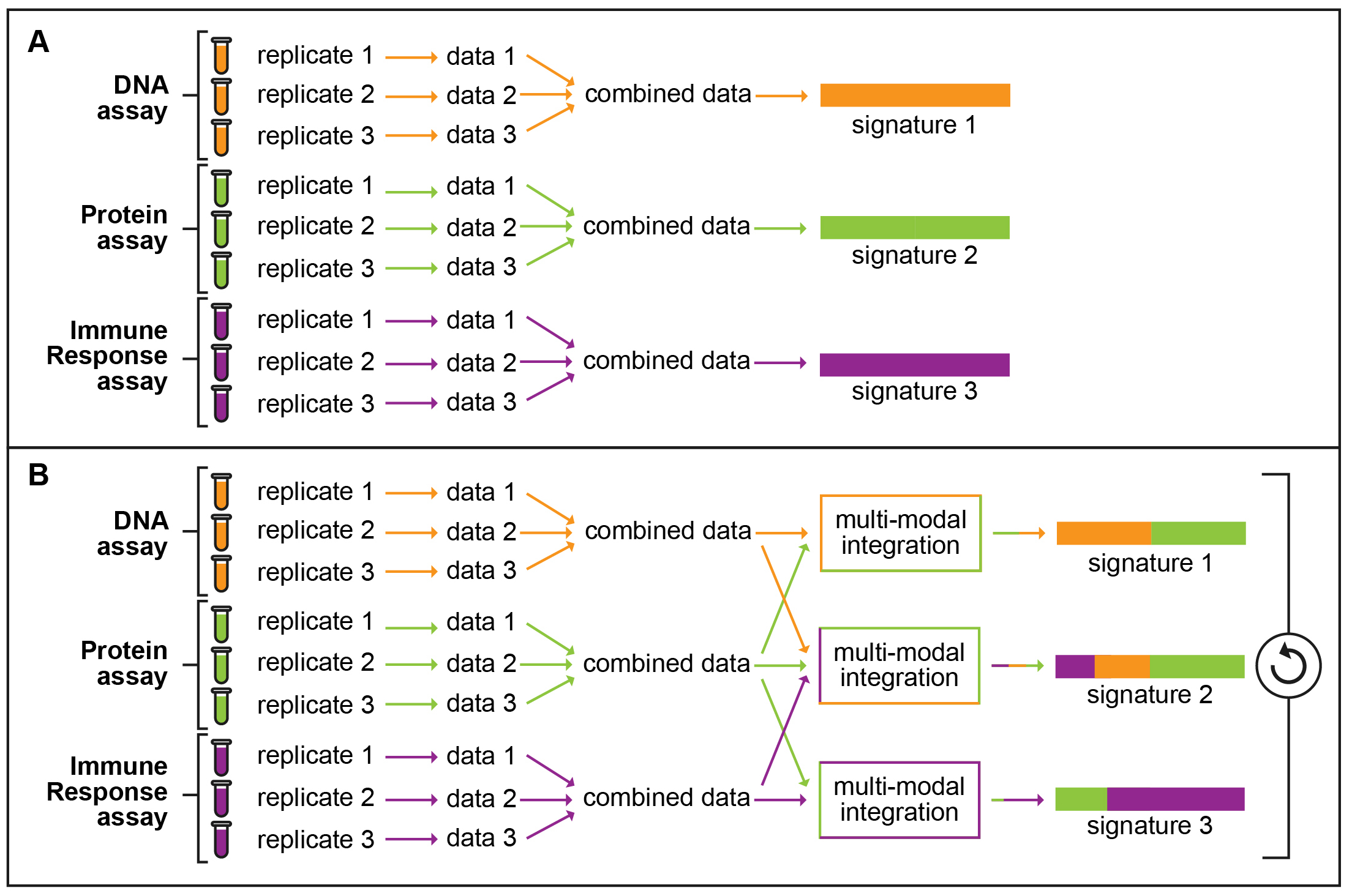} 
    \end{tabular}
    \caption{\textbf{Comparison of traditional versus a BAS-style approach toward signature development.} A) The traditional method for signature development. This method takes data from a single assay and generates a single signature from it. B) Proposed approach toward development of BAS signatures. In this method, data from multiple assays are combined together via multi-model integration to generate multiple signatures.
    }
    \label{fig:signature}
\end{figure}

As implied by the examples above, the technologies that are most useful in generating BASs will simultaneously measure the same sample using multiple platforms, because analyses that result from a single technology are unlikely to fully describe the various mechanisms underlying a response (Figure \ref{fig:signature}).
For instance, simply measuring inflammation can indicate a wide number of disease states ranging from autoimmune disorders to cancer to infection. Meta-analysis of these studies show diverse ranges of values, the variance in which could indicate subdomains with the overarching designation of \textit{inflammation} \cite{Coates2016,Furman2019,Hunter2012,Shibru2021}. Fully understanding these inflammation measurements necessitates the integration of immunological and multi-omics surveillance platforms. Defining and classifying what constitutes threat-induced immune responses will require tying `omics and host response data together to distinguish known from unknown. 

A comprehensive system for the detection of infection of humans would also require the simultaneous evaluation of foreign molecules and modulations in host responses. This system would enable both aspects of BAS signature development of pathogen and host response characterization.
A number of diagnostic methods in use today could be optimized to do this; for example, multianalyte LFIA's can monitor for infection in wounds by detecting interleukin-6 and pathogen DNA simultaneously, assessing both the host response and presence of a specific pathogenic agent \cite{Klebes2022}. In addition, microsphere-based flow cytometric techniques allow detection of multiple analytes within a single low volume sample \cite{Figueiredo2022}. As previously noted, it is now possible to measure mRNA and protein levels (via sequencing and mass spectrometry, respectively) from one single, unique cell \cite{Fulcher2022.05.17.492137}. However, one practical challenge rests in defining which analytes represent a host-dependent threat indicator in a way that is not agent specific. More broadly, as we will discuss later, analyzing data that have been jointly acquired from different modalities remains a challenge. 

\section{Common Gaps in Computational Biology}

\subsection{Absence of Baseline Information}
One major gap for signature development is the lack of a consistent baseline that would allow scientists to extrapolate signatures. In order to be capable of recognizing the presence of a threat, it is critical that we define composition of the \textit{normal system}.
For example, as of 2023 no public model of a healthy generic immune system exists to facilitate comparisons against data pertaining to disease states.
The non-systematic accumulation of patient samples over the decades comes primarily from individuals with a disease phenotype and gives wide ranges of values for individual measurements. The inference of disease state necessitates access to a healthy control group. This holds true whether we are comparing genetic data amongst known sequenced genomes, sifting through proteomics data from environmental samples, conducting metabolic analyses of human blood samples, or measuring cellular responses in patients. The successful development of BASs will likely require investments in healthy longitudinal cohorts to systematically probe healthy populations.

Although the community recognizes the importance of baselines and negative controls, few projects are dedicated to measuring these baselines.  Programs such as ``All in One Breath,'' which aims to understand what molecules people in a healthy population exhale, represent important progress, but only a few  publications have investigated similar themes\cite{pmid33710326}; additional work in this area is needed.

\subsection{Complete and Detailed Metadata Required for Signature Development}
Detailed metadata annotations of the collected data are necessary for BAS development. Because data production has outgrown any individual researcher's ability to analyze data manually, scientists increasingly use algorithms to elucidate  biological signals. Obtaining and identifying useable metadata are therefore increasingly important to ensure analyses produce sensible outputs. Standards such as the FAIR Principles (Findability, Accessibility, Interoperability, and Reuse of digital assets) aim to normalize metadata annotation for reuse \cite{wilkinson:fair}. Unfortunately, while data is routinely shared, data reuse is still remains a challenge due to barriers such as unstandardized metadata collection and uncoordinated data dissemination \cite{hughes:addressing}. Policies that promote data sharing, such as those the NIH implemented 2023 \cite{rothstein:informed,hughes:addressing}, support broader adoption of embracing FAIR Principles. If widely embraced, open sharing will unleash a wealth of data supporting the development of BASs.

Identifying BASs necessitates a metadata analytical approach, but no standardization exists that allow metadata to move beyond its association with an individual dataset or analysis. Precision and consistency across studies are required in metadata to prevent masking important biological differences. For example, although \textit{prostate cancer} and \textit{prostate cancer free} are valid metadata annotations, they are not precise because cancer subtypes may have different mechanisms and treatments. In addition, associating metadata with individual experiments, rather than with entire conclusions, supports better future use. For example, metadata on PRIDE \cite{martens:pride,vizcaino:update}, a public proteomics repository, is associated with entire reports rather than individual runs. This prevents future analysis of the dataset if the file naming scheme is unclear, in turn inhibiting the development of BASs.

Finally, different metadata schemes must be seamlessly interchangeable. Currently, each field has their own metadata scheme which does not necessarily relate to the schemes of other fields. The lack of a common standard to discuss and relate datasets hinders multidisciplinary research. Additional complexity arises from the different identifiers used across different nomenclature systems. For example, the gene and protein IDs for the same entity differ across systems such as NCBI, Uniprot, and EBI. While these entities offer the ability to translate across nomenclature systems, using them, in practice, remains challenging. For example, the mapping across nomenclature systems can be imprecise due to many-to-many relationships (e.g., a gene can be affiliated with multiple gene products).
In addition, another challenge is that these translation systems have to be constantly updated as the entries in different nomenclature systems are added or updated.
Creating consistent and widely used mappings between different metadata schemes would greatly enable multidisciplinary, integrative research that reflects the nature of BAS characterization and identification. 

\subsection{Data Harmonization}
Multi-omic analysis necessitates our ability to integrate disparate datasets for a unified analysis. Data integration is simplified if the experiments were designed for such an analysis \textit{a priori} so as to build anchor points into each independent data subset. Historically, studies each tested a single hypothesis without consideration for future use. The emergence of public data repositories in the past several decades spurred the development of integrative methods to analyze archived data for additional biological insight \cite{pmid27555308,pmid36472572,pmid35701420,pmid32695065}. Although there has been some success, additional investment is needed to further improve these computational techniques. 

\begin{figure}
    \centering
    \includegraphics[width=10cm]{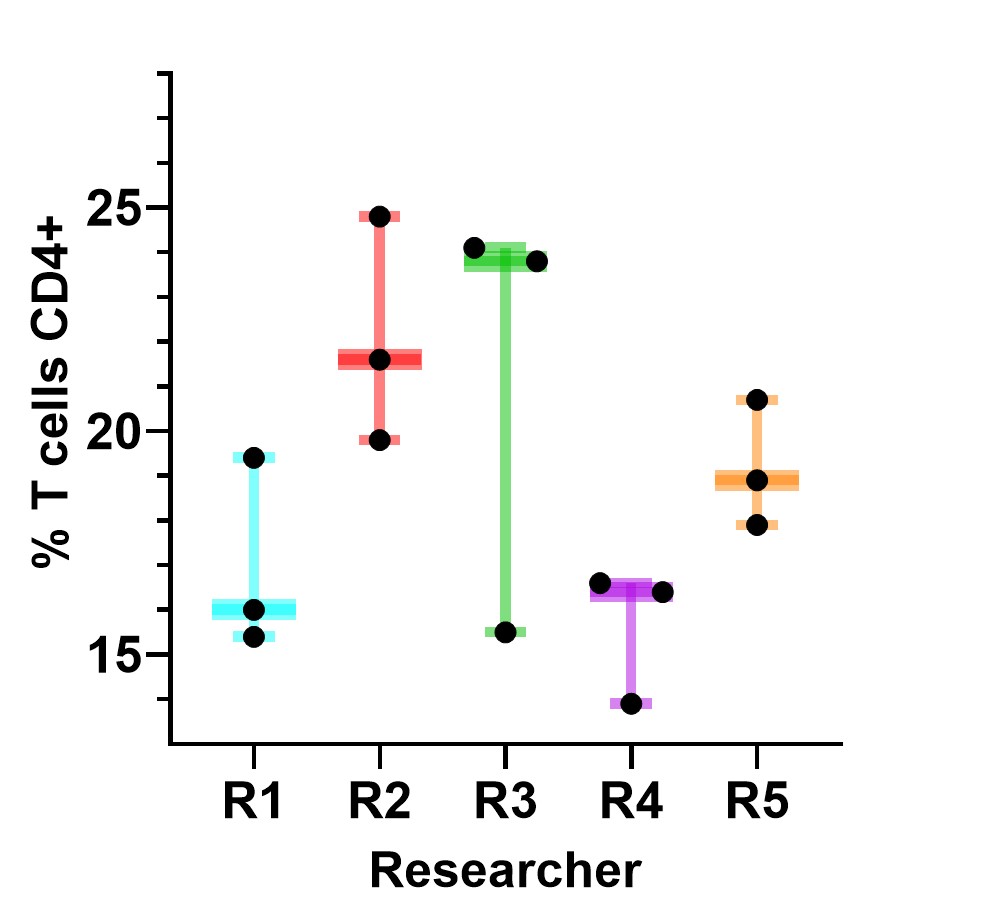}
    \caption{\textbf{Experimental batch effect hinders data harmonization.} Simultaneous triplicate flow cytometric analysis of the percentage of CD4+ T cells by five separate researchers (R1-R5) of an identical sample acquired on a Beckman Coulter Gallios flow cytometer. There is high variability of the measured percentage of CD4+ T cells within and between researchers. This variation is an example of the difficulty in determining typical vs atypical results for even well-established current assays that have been in use for decades. Data in this Figure are from the laboratory of C Spencer.}
    \label{fig:batchEffect}
\end{figure}

Analyzing data across multiple experiments requires careful thought due to issues associated with batch correction and normalization. Current bench science limitations require that data be generated in multiple batches, rather than all at a single time (the additional redundancy is an incidental benefit). Batches have both within and between systematic heterogeneity (Figure~\ref{fig:batchEffect}), otherwise known as batch effect, that are confounding. Samples (even within batches) need to be normalized to account for heterogeneity that can arise from the sample-specific treatments essential to the study whereas between batch heterogeneity often arises from systematic effects due to experiment external factors. 
These differences must be controlled/mitigated when analyzing data across batches. While multiple strategies exist for batch correction \cite{phua:perspective, uklina:diagnostics, leek:tackling, tran:benchmark} and batch normalization \cite{pmid20979621,pmid20196867},  
these processes are labor intensive, often requiring manual oversight. Future work is needed to improve and standardize batch correction and normalization methods when combining data from the same technology (e.g., flow cytometry), different technologies functionally measuring the same event (e.g., microarrays and next-generation sequencing), and different technologies measuring different yet correlated events (e.g., next-generation sequencing and metabolomics). Although sequence-based count batch effect is discrete, continuous batch effects do occur (e.g., when measuring an analyte on a mass spectrometer, its $m/z$ value will drift over time) \cite{phua:perspective,uklina:diagnostics}. As a result, methods that can simultaneously correct both discrete and continuous batch effects are also needed for multi-modal integration.

Successful batch correction necessitates detailed metadata and sound experimental design to allow correct stratification and prevent confounding variables from being introduced. For example, if cancer samples were sequenced at one facility and control samples were sequenced at a second facility, it would become impossible to disentangle the cancer signal from the processing facility if proper control of confounders had not been planned for from the beginning. Proper experimental design prior to data generation can minimize these sampling problems and provide the details needed to successfully integrate experiments into a unified meta-analyses. 

\subsection{Scaleable Computational Models are Needed for Data-Centric Continual Integration}
Exponential increases in data-acquisition rates necessitate new computational platforms and methods.  Researchers are today already generating more data than we have the computational capacity to analyze.  By 2025, an estimated 2--40 exabytes (millions of terabytes) will be needed just to store human genomes \cite{stephens:big}. Beyond the mere cost considerations, current models cannot integrate data at this scale, requiring both advanced storage abilities and novel algorithmic development \cite{muir:real}.

Machine learning (ML) models have become an increasingly popular tool for analyzing data, despite their need for advanced computational resources. 
ML models thrive when given 1. large, extensive, curated training data; 2. complex architectures enabling model expressivity (i.e., the ability of a model to estimate increasingly complex functions); and 3. the requisite hardware to fit these models. State-of-the-art models now require orders-of-magnitude more training using hundreds/thousands of graphical processing units (GPUs) at significant cost (compute resource/money/energy). ML scientists focused on advancing capabilities have largely ignored these costs, but searching for BASs will require more resource-efficient models that can yield high precision and recall in analyzing continuous streams of biological data.  

Developing a BAS is, itself, insufficient to meaningfully improve the state of public health and biodefense.  Operationalizing a BAS requires additional work, often involving adapting methods to reflect the realities of data acquisition, movement, and processing in real-world settings. For example, methods developed in a research setting generally have simultaneous access to all the data generated for a study. Such methods require adaptation for situations in which data is continuously generated in real time, such as during clinical bio-surveillance or environmental sampling. Computational methods will therefore need to be developed that can continuously analyze streaming data. In addition, these methods must be computationally efficient and ideally require minimal computational resources. This is not yet possible with the current implementation of statistical models. 

\section{Improved Omics Analyses for Systems Biology are Necessary}

\subsection{Standard Analyses of Omics Data are not Agent-Agnostic}
Although omics technologies, such as genomics and proteomics, have revolutionized our ability to detect pathogens and the immunological responses to them, the data that result from these technologies are typically analyzed in an agent-specific manner. For example, short-read sequencing is currently one of the most dominant genomic technologies used in the public health and biodefense fields, and the resulting reads are almost always aligned against reference genomes. 

Without reference genomes, we must assemble the sequences \textit{de novo}, which is a formidable task given the standard short-reads from sequencers; our limitation for \textit{de novo} assembly is exponentially increased for environmental samples. As a result, pathogen detection in environmental samples is performed using PCR. PCR is not an agent-agnostic assay, however, because primers (i.e. known targets) need to be constructed for use during amplification. 

Analysis of data from other 'omics technologies, such as proteomics and metabolomics, suffers from the same challenge. The standard proteomics data analysis involves detecting peptides in a sample by comparing experimentally derived spectra against a database of theoretically derived spectra \cite{eng:face}. These theoretically derived spectra are generated from peptide sequences, which are themselves derived from a reference genome. Of course, if one knows the identity of a sample, then a reference-based approach yields better results. In the case of metabolomics, untargeted metabolite identification usually requires the use of spectral libraries, which are typically generated from analyzing synthesized chemical standards \cite{bittremieux:critical,neumann:computational}.

\subsection{Continued Development and Adoption of \textit{de novo} Methods are Needed}
Creating BASs will require developing and adopting {\it de novo} algorithms for extremely complex samples. 
For example, long-read sequencing has the potential to greatly facilitate \textit{de novo} genome and transcriptome assembly of pathogens and hosts. Although assembling genomes solely on short sequence reads is generally feasible\cite{paszkiewicz:denovo}, it fails on repetitive regions \cite{treangen:repetitive} and complex metagenomic datasets \cite{ayling:new,marx:method}. Long-read sequencing technologies overcome these challenges, resulting in the completion of the human reference genome \cite{nurk:complete} and low-complexity microbiomes \cite{somerville:long}. These technologies have been embraced by the academic community, but have yet to be widely used in the biodefense or public health spheres. Improvements to decrease error rates \cite{amarasinghe:opportunities} are still needed, as is fundamental research into characteristic differences between long-read transcriptomics data and long-read genomics data \cite{amarasinghe:opportunities}. Furthermore, additional development is required in order to allow usage in complex microbial environmental monitoring.

Similarly, advances in \textit{de novo} analysis of proteomics data are necessary before integration in public health or biodefense. Although \textit{de novo} peptide detection algorithms exist \cite{tran:deep,tran:denovo}, additional work is needed to increase analytic power, increase speed of data collection and analysis, decrease error rates, and increase utility with respect to reference-based methods \cite{muth:evaluating}. In addition, challenges remain with detection and localization of novel and rare post-translational modifications (PTMs) \cite{miller:overview}. The field has almost exclusively focused on phosphoproteomics at the expense of other PTMs such as glycosylation, which poses a particular challenge due to the diversity of their complex structures \cite{grabarics:mass}. The ability to identify glycosylated biomolecules is important because they have been strongly implicated in host-pathogen interactions \cite{lin:cells,qin:host} and because 60\% of proteins are possibly glycosylated \cite{timp:beyond}. Finally, as new non-mass spectrometry-based technologies are developed, such as single-molecule protein sequencing \cite{yu:unidirectional,alfaro:emerging,timp:beyond}, new computational methods will be needed to analyze the resulting data with their unique biases, assumptions, strengths, and weaknesses, as well as to integrate the data with that of other technologies, as described above.

The usage of metabolomics in identifying BASs suffers from several computational challenges. The majority of untargeted metabolite identification uses spectral library searching of liquid chromatography-tandem mass spectrometry data \cite{bittremieux:critical,neumann:computational}. This method is insufficient because it requires previous analysis of an extensive set of chemical standards, estimated to exceed the number of atoms in the universe \cite{lemonick}. This necessitates the improvement of current methods, which currently have poor performance and return numerous candidates, that can predict structures, spectra, and functions of completely novel chemicals\cite{hrkop:searching,stravs:msnovelist,vanderhooft:topic}.  

\subsection{Improvements for Integrating Multi-modal Data are Needed}
\label{sec:multi}
BAS development will likely rely on signatures that integrate multiple data sources to obtain a signal that is more robust to noise and error. Using multiple, different assays can also detect complementary signals that a single assay would be unable to detect. For example, genomics can describe an organism's potential processes, but cannot inform which are active; conversely, proteomics can determine active processes but is unable to contextualize. Together, we obtain a more holistic understanding of the presence of a pathogen and the host response. 

Developing multi-modal signatures is not trivial \cite{tarazona:undisclosed}, requiring extensive interdisciplinary research and coordination among researchers and funding agencies. Common challenges include determining metadata sets and identifiers, and functional linking of different schemes. In addition, merging data from different assays is challenging because each assay has differing biases, variances, and power. For example, proteomics data acquired via data-dependent acquisition has missing data that correlates with abundance \cite{matafora:missing}, while transcriptomics data has missing data that correlates with time\cite{pmid30192227}. Furthermore, there are many different integration strategies \cite{canzler:prospects}, necessitating generalized integration methods that can be applied to a large number of assays.

The recent growth of single-cell `omics technology provides new tools enabling multi-modal signatures \cite{Fulcher2022.05.17.492137,pmid33835024,pmid31270458}, particularly because such methods may mitigate the challenges of computationally integrating disparate datasets across assays. The technology is still immature, however, and realizing its potential will require investments to understand how biases may affect correlations between modes. Until we understand these effects, their utility is limited in the search for immunological BASs.

The interrogation and analysis of biological pathways is another promising avenue for generating BASs. Fully understanding a pathway requires a systems biology approach that is most successful when multiple data streams, such as genomics, metabolomics, proteomics, and glycomics, are integrated for a single purpose, e.g., immunomics. Multi-modal analyses has the potential to generate signatures based off a small number of pathways related to pathogenicity and immune response against potential biothreats, especially if we can leverage single-cell technologies. Additional work is required to improve multi-modal analysis before pathway-based approaches can be successful. 

\section{Discussion}
Although there is momentum to embrace bioagent-agnostic biosurveillance, for instance, by applying `omics monitoring of wastewater~\cite{CUI2021116415, doi:10.1021/es502546t}, the utility and impact of these technologies will be limited by the gaps in current computational and systems biology. Assays and technologies are starting to be developed to generate BASs, but computational analyses and tools need further investment, as delinated here. We advocate for converging the technology, the data, and the modeling as a unified goal enabling BAS identification, characterization, and ultimately detection. From the computational side, the community must standardize data collection and annotation, increase research in scalable algorithms and methods development (especially in data harmonization for multi-modal analysis), and implement tools whose computational throughput can match the levels attained by new technologies such as single cell transcriptomics . Research in these disparate areas need to be unified, and dedicated investment is needed to convert these basic research areas into actionable capabilities.

Here we have highlighted the human immune system as a potential sensor for a threat agnostic biosurveillance system, but other sample types can also potentially act as sensors. For example, inter-kingdom interactions between bacteria, archaea, and fungi in environmental samples may yield signals related to antibiotic resistance~\cite{Schmidt2015}. In addition, although we have focused on human pathogens, this approach is generalizable and can address threats to crops, livestock, and the environment writ large. 
%Research in these disparate areas need to be unified, and the US government (via Defense, Health, Energy, Agriculture, etc.) should lead in the efforts to convert these basic research areas into actionable capabilities by identifying the most fruitful approaches. 

One important aspect to note is that BASs will evolve, and so we emphasize the need to update these markers through a data-centric continual integration approach. Our computational models will need to be refactored as pathogens evolve. The evolution of pathogens underscores the need for bioagent-agnostic signatures, which permits a flexibility in pattern recognition rather than focusing on a predetermined set of pathogens. While a BAS-approach will not be the final solution, a biothreat-agnostic approach will help close gaps that arise in list-based approaches.

\section{Acknowledgements}
Pacific Northwest National Laboratory is a multiprogram national laboratory operated by Battelle Memorial Institute for the United States Department of Energy under contract DE-AC06-76RLO. The views and conclusions contained in this document are those of the authors and should not be interpreted as necessarily representing the official policies, either expressed or implied, of the U.S. Government. We would like to acknowledge the assistance of Crystal Devora, Aylin Sanchez, Miguel Gutierrez and Luisa Rodriguez in generating data for this manuscript. This work was funded in part by National Institutes of Health award SC1AI148753.

\bibliographystyle{ama}
\bibliography{main} 
\end{document}